\documentclass[pre,nofootinbib,twocolumn,showpacs,preprintnumbers,amsmath,amssymb,floatfix,superscriptaddress]{revtex4}
\usepackage{amsmath}
\usepackage{amsfonts}
\usepackage{dcolumn}                    	     
\usepackage{amssymb}
\usepackage{graphicx,graphics,wrapfig,rotating}     	
\usepackage[english]{babel}	
\usepackage{bm,fancybox}                	     
\newcommand{\rmi}{{\rm i} }    
\newcommand{\rme}{{\rm e}}      

\begin{document}
\title{A periodic orbit basis for the Quantum Baker Map}
\author{Leonardo Ermann}
\email[Email address: ]{ermann@tandar.cnea.gov.ar}
\affiliation{Departamento de F\'\i sica, Comisi\'on Nacional de Energ\'\i a 
At\'omica, Avenida del Libertador 8250, (C1429BNP) Buenos Aires, Argentina}
\affiliation{Departamento de F\'\i sica, FCEyN, UBA, Pabell\'on 1 Ciudad 
Universitaria, C1428EGA Buenos Aires, Argentina}

\author{Marcos Saraceno} 
\affiliation{Departamento de F\'\i sica, Comisi\'on Nacional de Energ\'\i a 
At\'omica, Avenida del Libertador 8250, (C1429BNP) Buenos Aires, Argentina}
\affiliation{Escuela de Ciencia y Tecnolog\'\i a, Universidad Nacional de 
San Mart\'\i n, Alem 3901, B1653HIM Villa Ballester, Argentina}

\date{\today}

\begin{abstract}
A set of quantum states, dynamically related to the classical periodic orbits of a chaotic map, 
is used as a basis in which the description of the eigenstates of its quantum version is greatly simplified.
This set can be improved with the inclusion of short time propagation along the stable and unstable manifolds 
of the periodic orbits resulting in a construction similar to the scar functions of Vergini [Jour. Phys. A \textbf{33}, 4709 (2000)].
The average participation ratio is used to quantify the quality of the basis.
\end{abstract}

\maketitle

\section{Introduction}

The question of the structure of the eigenfunctions of chaotic systems is intimately related 
to the construction of the bases in which they can be expanded. In this respect a good basis is one
 in which most eigenstates are given in terms of a limited number of significant coefficients: 
a good measure being i.e. the average participation ratio in the given basis. In the generic 
case of a basis unrelated to the Hamiltonian (or the map) the states are mostly random and the 
average participation ratio ($\langle PR \rangle$) takes the random matrix value $D/2$ 
(for a Hilbert space of dimension $D$).
 At the other \emph{trivial} extreme the eigen basis gives a $\langle PR \rangle$ of unity. 
It is then clear that a ``good'' basis has to incorporate some dynamical elements from the system 
and at the same time be sufficiently simple so as to be effectively constructed without resorting 
to a full diagonalization.\\
In the case of the quantum baker's map (QBM) Lakshminarayan found that the 
eigenfunctions had a simple structure (and significantly small participation ratio) when looked upon
 in the Hadamard basis \cite{lakshmin}, thus exploiting a very special property of the QBM. This line of research 
was followed in \cite{leo2}, where it was found that the eigenfunctions of a large family 
of quantizations of the QBM could be described in terms of a very simple map, the essential baker,
 which for special values of the Hilbert space dimension becomes the Walsh quantized baker 
and can be explicitly constructed. These features 
are very special and intimately related to the binary symbolic dynamics of the map and are not easily 
generalized to other systems.\\
A different approach with a potentially more general applicability, is based on the construction of 
states that ``live'' on the unstable periodic orbits of the system. When used as a basis these states
realize in quantum mechanics the ideal of Poincar\'e in the sense that ``they (the periodic orbits) 
are the only breach through we might try to penetrate into a stronghold hitherto reputed 
unassailable'' \cite{poincare}.
The fact that some eigenstates of chaotic systems show ``scars'' of periodic orbits was 
established long ago by Heller
\cite{Heller},
 in counterpart to earlier works in which the assumption was
a uniform distribution on the energy shell according to the 
microcanonical ensemble \cite{Berryvoros,schnirelman}.
The scarring phenomena was studied in several chaotic systems, 
in which linear and non-linear theories were developed \cite{kaplan2,kaplan3,kaplan4,vergini,vergini2,creagh,nonnen}.\\
The construction of scar functions in the QBM 
 is based on early studies in the stadium 
billiard \cite{vergini,vergini2,verginiborondo}.  The QBM can be thought of as a pedagogical system 
to apply these techniques since it has symbolic dynamics, stable (unstable) manifold parallel to $q$ ($p$) 
direction, a finite spectrum and the same 
small-valued Lyapunov exponent in the entire phase space.\\
In this paper we provide a recipe to construct a set of scar functions 
as an accurate basis to describe the QBM. This basis has the 
propagation time of the quantum propagator as parameter. When this time is 
of the order of the Heisenberg time, the basis converges to the eigen base 
of the map. For short times, of the order of the Ehrenfest time, the basis 
describes the spectrum of the QBM better than other known bases \cite{lakshmin,leo2}.\\
In section II, we briefly introduce the classical and quantum 
version of the baker map. We construct the periodic orbit modes and scar functions based 
on the evolution of the coherent states under the QBM. 
We give the rule with which we choose a basis to describe the spectrum of 
the QBM for any even dimensional Hilbert space in section III. Then, in section IV, 
we numerically test the basis, computing the average participation ratio as a function 
of the propagation time.
In section V we propose a method to approximate the scar functions by homoclinic periodic 
orbit modes avoiding evolution in time, and finally, we state the conclusions.

\section{Classical and Quantum Evolution}

\subsection{The baker's transformation}

In this section we review some properties and notation of the classical baker's map that we will 
use in the quantum states construction. The baker's map $\mathcal{B}$ \cite{arnold} is defined in 
the unit square phase space ($q,p \in [0,1)$) as
\begin{eqnarray}
q^\prime&=&2q-\lfloor2q\rfloor  \nonumber\\
p^\prime&=&\frac{(p+\lfloor2q\rfloor)}{2}
\end{eqnarray}
where $\lfloor x\rfloor$ is the integer part of $x$.
This map is area-preserving, uniformly hyperbolic with Lyapunov exponent 
($\lambda=\ln{2}$), and has stable foliation $\left\lbrace q=\text{cst} \right\rbrace$ 
and unstable foliation $\left\lbrace p=\text{cst} \right\rbrace$.

The baker map has a simple action upon symbols in the binary expansion of the coordinates
\begin{equation}
(p|q)=\ldots\epsilon_{-1}\cdot\epsilon_{0} 
\epsilon_{1}\ldots
\stackrel{\mathcal{B}}{\longrightarrow}
(p\prime|q\prime)=\ldots\epsilon_{-1}\epsilon_{0}
\cdot\epsilon_{1}\ldots
\end{equation}
where $q=\sum_{i=0}^{\infty}\epsilon_{i} 2^{-(i+1)}$ and $p=\sum_{i=-1}^{-\infty}\epsilon_{i} 2^{i}$.

The map has two symmetries:
\begin{itemize}
\item Parity ($R$): represented with the exchanges \mbox{$q\rightarrow 1-q$}
 and $p\rightarrow 1-p$ together with the bitwise logical NOT upon
 symbols $(0\leftrightarrows1)$.
 \item Time reversal ($T$): represented with the exchange \mbox{$p\leftrightarrows q$}
 together with reversing the direction of the symbolic flow.
\end{itemize}

The periodic orbits of the baker map of period $L$ can be represented by binary strings $\bm{\nu}$ 
of length $L$.
We denote the different trajectory points on a periodic orbit by \mbox{$(q_j,p_j)$} for $j=0,\cdots,L-1$ 
with $(q_L,p_L)\equiv(q_0,p_0)$.
The coordinates of the first trajectory point on the periodic orbit can be 
obtained explicitly in terms of the binary string 
\begin{eqnarray}\label{eq:traj1}
q_0&=&\cdot\bm{\nu\nu\nu}\ldots=\frac{\nu}{2^{L}-1}\\ 
p_0&=&\cdot\bm{\nu^{\dagger}\nu^{\dagger}\nu^{\dagger}}
\ldots=\frac{\nu^\dagger}{2^{L}-1} \label{eq:traj2}
\end{eqnarray}
where $\nu$ is the integer value of the string $\bm{\nu}$ which represents a binary number, and 
$\bm{\nu^{\dagger}}$ is the string formed by all $L$ bits of 
$\bm{\nu}$ in reverse order.
The other trajectory points can be easily calculated by iterations of the map or by cyclic shifts of $\bm{\nu}$.

\subsection{The Quantum Baker Map}

The quantization of the map is performed in an even $D$-dimensional Hilbert space with $D=1/(2\pi\hbar)$. 
The QBM is defined in terms of the discrete Fourier 
transform with antisymmetric boundary conditions as \cite{voros,saraceno,saravoros}
\begin{eqnarray}
\hat{B}&=&\hat{G}_{D}^{\dagger}\left( \begin{array}{cc}
 \hat{G}_{D/2}&0\\0&\hat{G}_{D/2}
\end{array}
\right)\\
\langle j\vert\hat{G}_{D}\vert k\rangle&=& \frac{1}{\sqrt{D}}
\exp{\left\{-i\frac{2\pi}{D} \left(j+\frac{1}{2}\right)\left(k+\frac{1}{2}\right)\right\}}
\end{eqnarray}

The quantum baker map has the same symmetries as its classical counterpart
\begin{eqnarray}
\left[\hat{B},\hat{R}\right]&=&0\\
\left( \hat{G}\hat{B}\hat{G}^{-1} \right)^{*}&=&\hat{B}^{-1} 
\end{eqnarray}
with parity represented by $\hat{R}=-\hat{G}^{2}$ and time reversal by $\hat{T}=\hat{K}\hat{G}$, 
where $\hat{K}$ is the complex conjugation operator.

The QBM spectrum is characterized by $D$ eigenphases and eigenstates $\hat{B}\vert\psi_{j}\rangle=
e^{i\varphi_{j}}\vert\psi_{j}\rangle$, with definite $\hat{R}$ symmetry 
($\hat{R}\vert\psi_{j}\rangle=\pm\vert\psi_{j}\rangle$), and satisfying the time reversal requirement $\hat{G}\vert\psi_i\rangle=\vert\psi_i\rangle^*$.

\section{Basis construction}
\subsection{Periodic Orbit Modes}\label{sec:POM}

The first step in our construction is the definition of the periodic orbit modes (POM), a 
superposition of coherent states centered on the periodic points of an orbit. Similar constructions have 
been employed before \cite{simonotti,kaplan2,kaplan3,kaplan4,vergini,vergini2}. Our definition is equivalent to the 
discrete--time version of the tube functions defined by Vergini and Carlo in \cite{vergini,vergini2} for continuous 
Hamiltonian flows.

The coherent state on the $D$-dimensional Hilbert space on the torus with anti--periodic boundary conditions 
centered on $(q,p)$ (\cite{coherent}) are represented in coordinate basis $\vert j\rangle$ as 
\begin{equation}
\langle j\vert q,p \rangle=K\sum_{m=-\infty}^{\infty}\rme^{-\pi D
\left( e_{j}+m-q \right)^{2}}
 \rme^{\rmi 2\pi D \left(e_{j}+m-q/2 \right)p-\rmi\pi m}
\end{equation} 
where $e_{j}=(j+1/2)/D$ and $K$ is a normalization factor which converges to $(2/D)^{1/4}$ for $D\gg1$. 
The phase has been chosen in such a way that parity and time reversal operators  act on them as
\begin{eqnarray}
\hat{R}\vert  q,p \rangle&=&\vert 1-q,1-p \rangle\\
\hat{T}\vert  q,p \rangle&=&\vert p,q \rangle
\end{eqnarray}
without additional phases.

\begin{figure}[htpb]
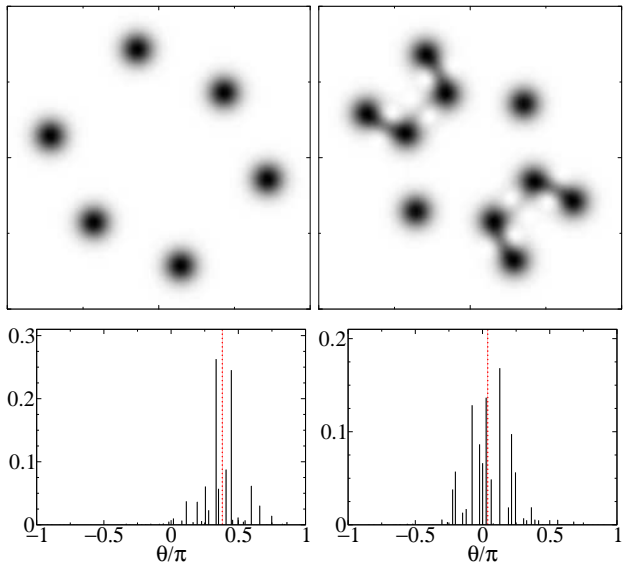

\begin{center}
 \includegraphics[width=0.225\textwidth]{1a.ps}
  \includegraphics[width=0.225\textwidth]{1b.ps}\\
\vspace{0.2cm}  \includegraphics[width=0.225\textwidth]{1c.eps}
  \includegraphics[width=0.225\textwidth]{1d.eps}
\caption{(color online) Top: Husimi representation of the periodic orbit modes over $001$
$\vert\tilde{\Phi}_{001}^{1,R=-}\rangle$ (left) and over 
$00101$ $\vert\tilde{\Phi}_{00101}^{3,R=+}\rangle$ for $D=120$ in the unit square phase space $q,p\in[0,1)$.
Bottom: Square modulus of the product of these states $\vert\langle\psi_{j}\vert\tilde{\Phi}_{001}^{1,R=-}\rangle\vert^{2}$ (left) and $\vert\langle\psi_{j}\vert\tilde{\Phi}_{00101}^{3,R=+}\rangle\vert^{2}$
with eigenstates $\vert\psi_{j}\rangle$ with $j=1,\ldots,D$ 
ordered by increasing eigenphases. The red dashed line represents the central phase $A_{001}^{k=1}$ and $A_{00101}^{k=3}$.}\label{fig:POM}
\end{center}
\end{figure} 

We now consider the collection of coherent states on the periodic points $\vert q_i,p_i\rangle$, 
$i=0,\cdots,L-1$ of a given primitive orbit labeled by the binary string $\bm{\nu}$. 
For chaotic systems the points are isolated and therefore in the semiclassical limit $D\to\infty$ 
these states are approximately orthogonal. In the same semiclassical limit they satisfy the 
approximate conditions
\begin{eqnarray}\label{eq:matrix}
\langle q_{j+1},p_{j+1}\vert q_{j},p_{j}\rangle &\simeq& \delta_{j+1,j}\\ \nonumber
\hat{B}_{j\ j+1}\equiv\langle q_{j+1},p_{j+1}\vert\hat{B}\vert q_{j},p_{j}\rangle &\simeq& \frac{\rme^{\rmi2\pi D S_{j}}}{\sqrt{\cosh{\lambda}}}
\end{eqnarray}
where $q_L\equiv q_0$, $p_L\equiv p_0$; $\lambda$ is the Lyapunov exponent and where the phase 
$S_{j}$ acquired by the coherent state in one step of the map (with the present choice of phases for 
the coherent states) is
\begin{equation}
S_j\equiv S_{q_j,p_j}=\left[ 2q_j\right]\left( \frac{q_j}{2}+\frac{p_j}{4}+\frac{1}{4}\right) 
\end{equation} 
The $L\times L$ matrix $\hat{B}_{j\ k}$ in Eq. \ref{eq:matrix} is cyclic in the semiclassical limit and therefore it 
can be diagonalized by a discrete Fourier transform. The eigenvalues are given by
\begin{equation}\label{eq:BSphase}
\langle\phi_{\bm{\nu}}^{k}\vert\hat{B}\vert\phi_{\bm{\nu}}^{k}\rangle 
   \simeq \frac{\rme^{\rmi2\pi A^{k}_{\bm{\nu}}}}{\sqrt{\cosh{\lambda}}}
\end{equation}
where
\begin{equation}\label{eq:Adef}
A^{k}_{\nu}=\frac{ D S_{\bm{\nu}}+k}{L}.
\end{equation}
The phase of the eigenvalues involves the classical action of the orbit
$S_{\bm{\nu}}=\sum_{j=0}^{L-1} S_{j}$, and $k$ is a Bohr-Sommerfeld like parameter which can be chosen from $k=0,\ldots,L-1$. 
Each periodic orbit then contributes $L$ complex eigenvalues whose phases are equally spaced and shifted from the 
origin by $D S_{\bm{\nu}}/L$.

The eigenfunctions are the periodic orbit modes. They are given explicitly by 
\begin{equation}\label{eq:POM}
 \vert\phi_{\bm{\nu}}^{k}\rangle=\frac{1}{\sqrt{L}}
\sum_{j=0}^{L-1}\exp\left(-\rmi\frac{2\pi (DS_{\nu}+k)j}{L}+\rmi\theta_j\right)\vert q_j,p_j\rangle 
\end{equation}
where $\theta_j=2\pi D \sum_{l=0}^{j-1} S_l$. They are labeled by the binary symbol of the periodic 
orbit and by the discrete index $k$ $(k=0\cdots L-1)$. Within the validity of the above approximations 
they are orthogonal. 
The fact that the eigenvalues are complex reflect the instability of the orbit and characterize 
these states as long lived resonances whose approximate width on the unit circle is 
$\lambda$. As this width is classical (independent of $D$) these resonances overlap significantly with a 
number of eigenstates $\lambda D/{2\pi}$ which is large in the semiclassical limit.  

It is convenient to impose the map symmetries to the POM's. 
The symmetries of the periodic orbits can be used to this purpose. 
The PO of the classical baker map can be classified in terms 
of their invariance under the classical symmetries $R$ and $T$. We characterize this invariance by two integers 
$\sigma_R, \sigma_T$ with the value $\sigma_{R}=0$ or $\sigma_{T}=0$ for invariant orbits, while $\sigma_{R}=1$ or $\sigma_{T}=1$ if there are two different 
orbits connected by the respective symmetry. As the action $S_{\bm{\nu}}$ is invariant under these symmetries, 
the eigenvalues of the POM constructed for each $S_{\bm{\nu}}$ are 
degenerate with an associated subspace of dimension $2^{\sigma_{T}}2^{\sigma_{R}}$. 
In these subspaces it is possible to construct POM's that have the same symmetries as the eigenfunctions. 
Thus central orbits ($\sigma_R=\sigma_T=0$) with $\bm{\nu}=\bm{\nu}^\dagger=\overline{\bm{\nu}}$ give rise to $L$ 
states, which automatically have the required symmetries. Orbits with either $\sigma_R=0$, $\sigma_T=1$ or $\sigma_R=1$, 
$\sigma_T=0$ give rise to $2L$ states, while non-symmetric orbits $\sigma_R=1$, $\sigma_T=1$ give rise to $4L$ states. 
Some examples illustrating this construction are 
\begin{eqnarray}
\vert\tilde{\Phi}_{01}^{k}\rangle&\equiv&\vert\phi_{01}^{k}\rangle \\
\vert\tilde{\Phi}_{001}^{k,R=\pm}\rangle&\equiv&\frac{\left( 
1\pm R\right)}{\sqrt{2}}\vert\phi_{001}^{k}\rangle \\
\vert\tilde{\Phi}_{001011}^{k,T=\pm}\rangle&\equiv&\frac{\left( 
1\pm T\right)}{\sqrt{2}}\vert\phi_{001011}^{k}\rangle \\
\vert\tilde{\Phi}_{0001011}^{k,T=\pm,R=\pm}\rangle&\equiv&
\frac{\left( 1\pm T\right)}{\sqrt{2}}\frac{\left( 1\pm R\right)}{\sqrt{2}} 
\vert\phi_{0001011}^{k}\rangle 
\end{eqnarray}
where $k=0,\ldots,L-1$.

\begin{figure}[htpb]
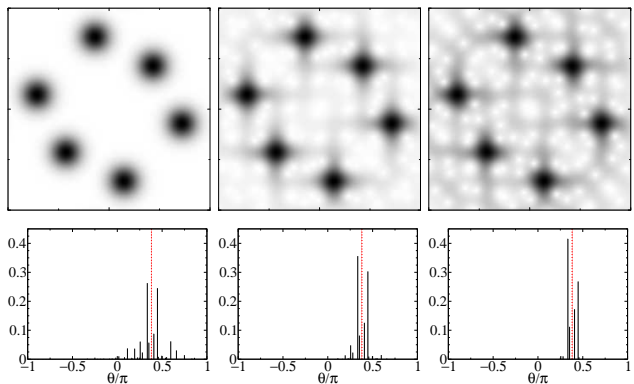

\begin{center}
 \includegraphics[width=0.15\textwidth]{2a.ps}
 \includegraphics[width=0.15\textwidth]{2b.ps}
 \includegraphics[width=0.15\textwidth]{2c.ps}\\
\vspace{0.2cm}  \includegraphics[width=0.15\textwidth]{2d.eps}
 \includegraphics[width=0.15\textwidth]{2e.eps}
 \includegraphics[width=0.15\textwidth]{2f.eps}
\caption{(color online) top: Husimi representation of the scar functions, in the unit square phase space ($q,p\in[0,1)$), over $\bm{\nu}=001$ with different propagation times
$\vert\Phi_{001}^{t=0,k=1,R=-}\rangle$ (t=0) (left), $\vert\Phi_{001}^{t=2,k=1,R=-}\rangle$ (t=2) (center) and $\vert\Phi_{001}^{t=4,k=1,R=-}\rangle$ (t=4) (right) for $D=120$. 
We plot the square of the Husimi function to enhance the structure on the manifolds.
bottom: Square modulus of the product of these states $\vert\langle\psi_{j}\vert\Phi_{001}^{t,k=1,R=-}\rangle\vert^{2}$ with eigenstates $\vert\psi_{j}\rangle$ with $j=1,\ldots,D$ 
ordered by increasing eigenphases. 
The red dashed line represents the central phase $A_{001}^{k=0}$}\label{fig:scar}
\end{center}
\end{figure} 

Figure \ref{fig:POM} shows the Husimi representation of two symmetrized POM corresponding to $\bm{\nu}=001$ and $\bm{\nu}=00101$.
The bottom part shows the distribution of the squared overlaps with the eigenstates as a function of the eigenphases. 
The central dotted line is the Bohr--Sommerfeld energy in Eq. \ref{eq:Adef}.
It should be clear that this construction can be justified for a fixed periodic orbit,
 and for $D\rightarrow\infty$ because the periodic points of chaotic systems are isolated. 
However if we want these quasimodes as a basis for a fixed value of $D$ we need also to consider 
orbits where the assumptions in Eq. \ref{eq:matrix} are not satisfied. Instead of a diagonalization by 
means of an explicit Fourier transform we have to consider the generalized eigenvalue problem 
$\det\left[\langle q_i,p_i\vert\hat{B}\vert q_j,p_j\rangle-\lambda\langle q_i,p_i\vert q_j.p_j\rangle\right]=0$. 
We explore this problem in connection to homoclinic orbits in the Appendix A.

\subsection{Scar functions}

In the previous section we have seen that the POM are quasienergy wavepackets of constant classical width $\lambda$.
 Narrower wavepackets can be constructed by Fourier transforming the POM's evolved for a limited time,
 \cite{vergini,vergini2,kaplan2,kaplan3,kaplan4,nonnen}. The resulting states are the scar functions.

Consider the following operator which depends on the time parameter $t$ and the phase $\varepsilon$
\begin{equation}
\hat{P}_{t}(\varepsilon)=\sum_{l=-\infty}^{\infty}e^{-i\varepsilon l}
e^{-\frac{l^{2}}{2t^{2}}} \hat{B}^{l}
\end{equation}
When the gaussian window is allowed to have infinite width ($t\to\infty$), this operator projects on the quasienergy eigenstates while for finite $t$ the corresponding width will be 
$\Delta\epsilon=2\pi/t$ 
In general we have
\begin{eqnarray}
\hat{P}_{t}(\varepsilon)&=
&\sum_{j=1}^{D}\sum_{l=-\infty}^{\infty}e^{-i\varepsilon l}e^{-\frac{l^{2}}{2t^{2}}}\langle\psi_{j}\vert 
\hat{B}^{l}\vert\psi_{j}\rangle \vert\psi_{j}\rangle\langle\psi_{j}\vert\nonumber\\
&=&\sum_{j=1}^{D}\delta_t(\varepsilon-\varphi_{j})\vert\psi_{j}\rangle\langle\psi_{j}\vert
\end{eqnarray}
where $\delta_t(\varepsilon)$ is the Fourier transform of $e^{-\frac{l^{2}}{2t^{2}}}$. 
When this operator acts on a POM it sharpens its quasi--energy width while extending the wave packet in phase space
along the stable and unstable manifolds of the periodic orbit. 
A wave function is thus created that interpolates between a simply constructed but relatively unstable state
and a true eigenstate if the propagation time is of the order of the Heisenberg time.
The interesting region is of course when the propagation time is of the order of the Ehrenfest time.
We define the {\em scar} functions as the short time propagation ($\propto$ Ehrenfest time $=\log_{2}{D}$) 
 of POM in $\bm{\nu}$ with the phase evaluated in $\varepsilon=A_{\bm{\nu}}^k$ defined at Eq. \ref{eq:Adef} 
\begin{equation}
\vert\Phi_{\bm{\nu}}^{t,k}\rangle\equiv\frac{1}{\kappa} \hat{P}_{t}\left(2\pi A_{\bm{\nu}}^k\right)\vert\tilde{\Phi}_{\bm{\nu}}^{k}\rangle
\end{equation}
where $\kappa$ is a normalization factor, and the POM is recovered for $t=0$ ($\vert\Phi_{\bm{\nu}}^{t=0,k}\rangle\equiv\vert\tilde{\Phi}_{\bm{\nu}}^{k}\rangle$). Notice that, for unstable periodic orbits, the forward and backwards propagation of wavepackets on the periodic points  lead to significant amplitude on the stable and unstable manifolds of that orbit. It is then 
expected \cite{vergini,vergini2} that 
these scar functions will be structures supported on these manifolds and having narrower overlaps with definite quasienergy regions on the unit circle.  
In this work we fix the phase $\varepsilon$ to the Bohr--Sommerfeld values ($A_{\bm{\nu}}^k$) and vary the time propagation $t$.

The Husimi representations of the scar functions over $\bm{\nu}=001$ for different evolution times $(t=0,2,4)$ , 
and their products with QBM eigenstates are shown in Fig. \ref{fig:scar}.
Note that the coherent states in the Husimi representation spread in the stable 
and unstable manifolds interfering between each other. Note that the Ehrenfest time for $D=120$ is approximately $t_{\text{Ehr}}\sim7$.

\section{Numerical test of the Basis}

\subsection{Scar Function Basis}

The QBM can be simplified using the scar functions as basis. 
In fact, it is usefull to choose a set of $D_{S}$ non--orthogonal 
scar functions, which span the $D$-dimensional Hilbert space, as an overcomplete basis of the QBM.
The election of the periodic orbits will determine the basis.
As we want to construct functions on short PO we give two different rules to choose the PO 
of the basis which converge to the same basis for long $D$. 

1) The first choice is to select the PO with shortest period which span the Hilbert space. 
For example, for $D=100$ we will have $D_S=106$ scar functions constructed over PO with 
period up to $L=6$, and for $D=140$ we have to include PO with period $L=7$ and therefore $D_S=232$ (see Table \ref{tab:PObase} for $D\leq472$).

2) The other possible election of the PO could be, 
for a given $D$, to choose $D_{S}=2^{L}\in[D,2D)$ and the 
short PO labeled by binary strings of $L$ digits. 

\begin{figure}[tpb]
\begin{center}
 \includegraphics[width=0.34\textwidth]{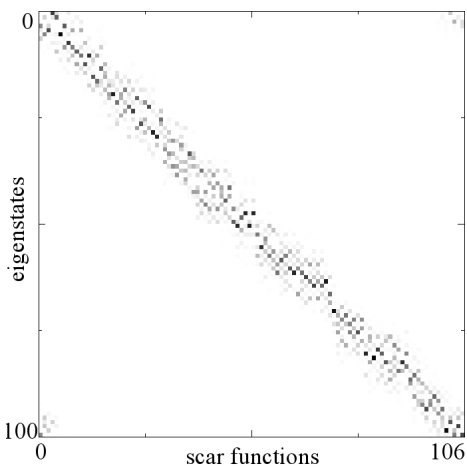}
\caption{Overlap matrix, $\vert\langle\psi_{j}\vert\Phi_{\bm{\nu}}^{t,k,R,T}\rangle\vert^2$, 
between the eigenstates of the QBM ordered by eigenphases and the scar 
functions constructed on the PO with $L\leq6$, ordered by increasing 
phase $A_{\bm{\nu}}^k$ for $D=100$ ($\vert\psi_{j}\rangle$ 
on rows and $\vert\Phi_{\bm{\nu}}^{t=t_\text{Ehr}/2,k,R,T}\rangle$ on columns). The value of the overlap is in grayscale from $0$ (white) to $1$ (black).}\label{fig:ProdscarD100}
\end{center}
\end{figure}

In this work we will choose the shortest period basis,
but for long $D$ limit both bases, are similar.
The important fact is that in both cases the maximum period involved grows as
$\log_{2}{D}$.
Fig. \ref{fig:ProdscarD100} shows the overlap matrix 
$\vert\langle\psi_{j}\vert\Phi_{\bm{\nu}}^{t,k,R,T}\rangle\vert^2$ of QBM eigenstates ($D=100$) ordered by eigenphases
($\vert\psi_{j}\rangle$, on rows) and the  scar function basis ($D_{S}=106$)
($\vert\Phi_{\bm{\nu}}^{t=t_\text{Ehr}/2,k,R,T}\rangle$, on columns) ordered by phase value $A^{k}_{\bm{\nu}}$, 
with $t=t_\text{Ehr}/2$.

\begin{table}[tp]%
\centering %
\begin{tabular}{lll}
\hline \hline%
dimension  & \multicolumn{2}{c}{Periodic orbit added}\\
$D_s$&$L_{\nu}$&$\nu$\\\hline
2&1& 0\\\hline
4&2& 01\\\hline
10&3 & 001\\\hline
22& 4& 0001; 0011\\\hline 
52& 5& 00001; 00011; 00101\\\hline 
106& 6& 000001; 000011;	 000101; 000111; 001011\\\hline 
232& 7& 0000001; 0000011; 0000101; 0001001;\\& & 0000111; 0001011; 0001101; 0010011; 0010101\\\hline
472 & 8 & 00000001; 00000011; 00000101; 00001001;\\ & & 00000111; 00001011; 00001101; 00010101;\\ & & 00011001; 00010011; 00100101; 00001111;\\ & & 00010111; 00011011; 00101011; 00101101\\\hline
\hline
\end{tabular}
\caption{\label{tab:PObase} Periodic Orbits used in basis construction of dimension $D_s$ to describe the QBM spectrum of dimension $D$ (where $D_s\ge D$).}
\end{table}


\begin{figure}[tpb]
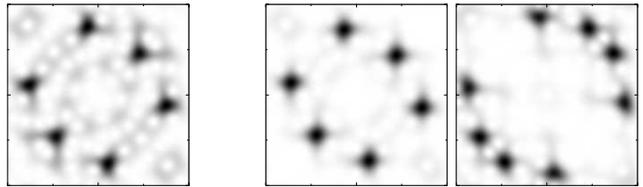

\begin{center}
 \includegraphics[width=0.135\textwidth]{4a.ps} \hspace{0.8cm}
 \includegraphics[width=0.135\textwidth]{4b.ps} 
 \includegraphics[width=0.135\textwidth]{4c.ps}
\caption{Husimi representation in the unit square phase space ($q,p\in[0,1)$) of an eigenstate of the QBM for $D=102$ (left), and its most prominent scar function components (right) as given in Eq. \ref{eq:ansatz}
}\label{fig:states}
\end{center}
\end{figure}

This result, besides restating the fact that scar functions have narrow overlaps with eigenstates also shows that an
eigenstate can be pictured as a narrow superposition of scars. Each of which is relatively simple to construct.
For example, in the case of $D=102$ one of the eigenstates of the QBM can be represented by only one scar function of period $L=3$ with a propagation time of $t=t_{\text{Ehr}}/2$ with an accuracy of $\vert\langle \psi_j\vert \Phi_{001}^{t_{\text{Ehr}/2},k=2,R=+}\rangle\vert\simeq0.87$. 
The approximation of the eigenstate can be improved adding another scar function of period $L=4$. 
Therefore an ansatz state constructed as 
\begin{equation}\label{eq:ansatz}
 \vert \Phi_{\text{ansatz}}\rangle=c_1\vert \Phi_{001}^{t_{\text{Ehr}}/2,k=2,R=+}\rangle+
c_2\vert \Phi_{0001}^{t_{\text{Ehr}}/2,k=2,R=+}\rangle 
\end{equation}
with $c_1\simeq0.75$ and $c_2\simeq0.48e^{-i0.117}$ has an accuracy of 
$\vert\langle \psi_j\vert \Phi_{\text{ansatz}}\rangle\vert\simeq0.92$. 
Figure \ref{fig:states} shows the husimi 
functions of the eigenstates and both scar functions considered in this example. 

\begin{figure}[tpb]
\begin{center}
 \includegraphics[width=0.43\textwidth]{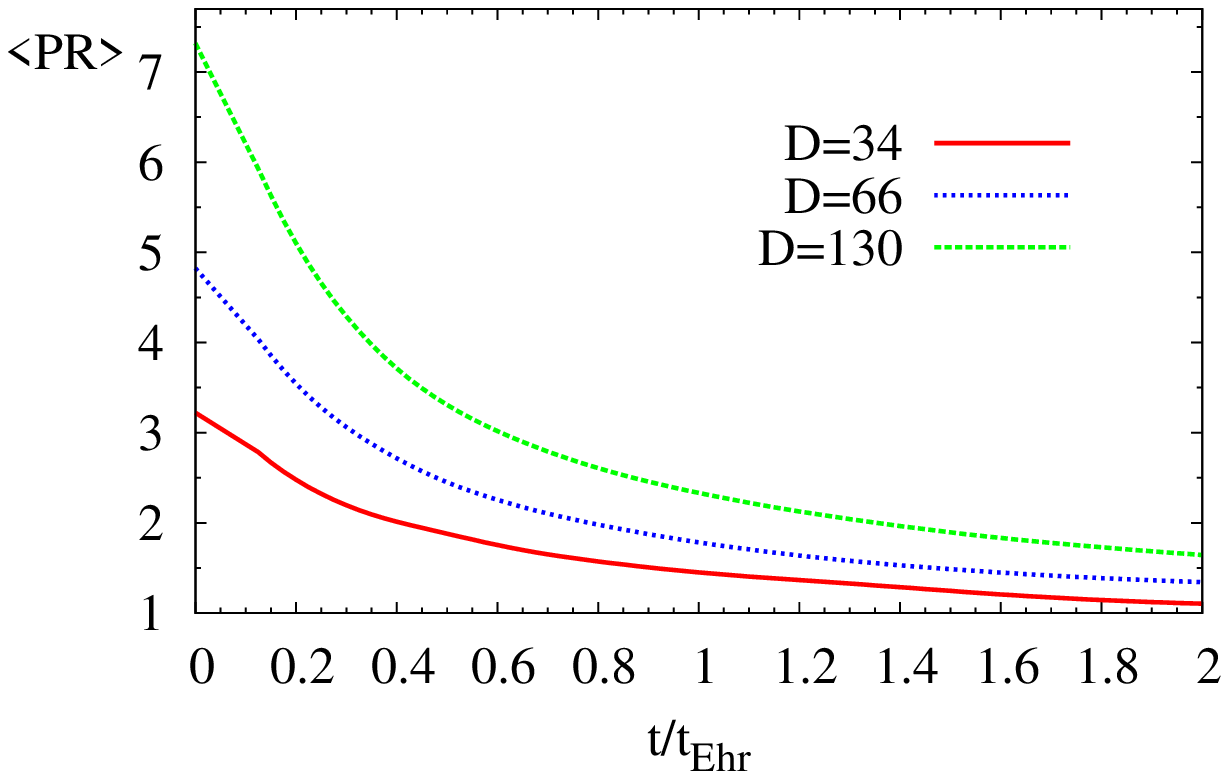}
 \includegraphics[width=0.43\textwidth]{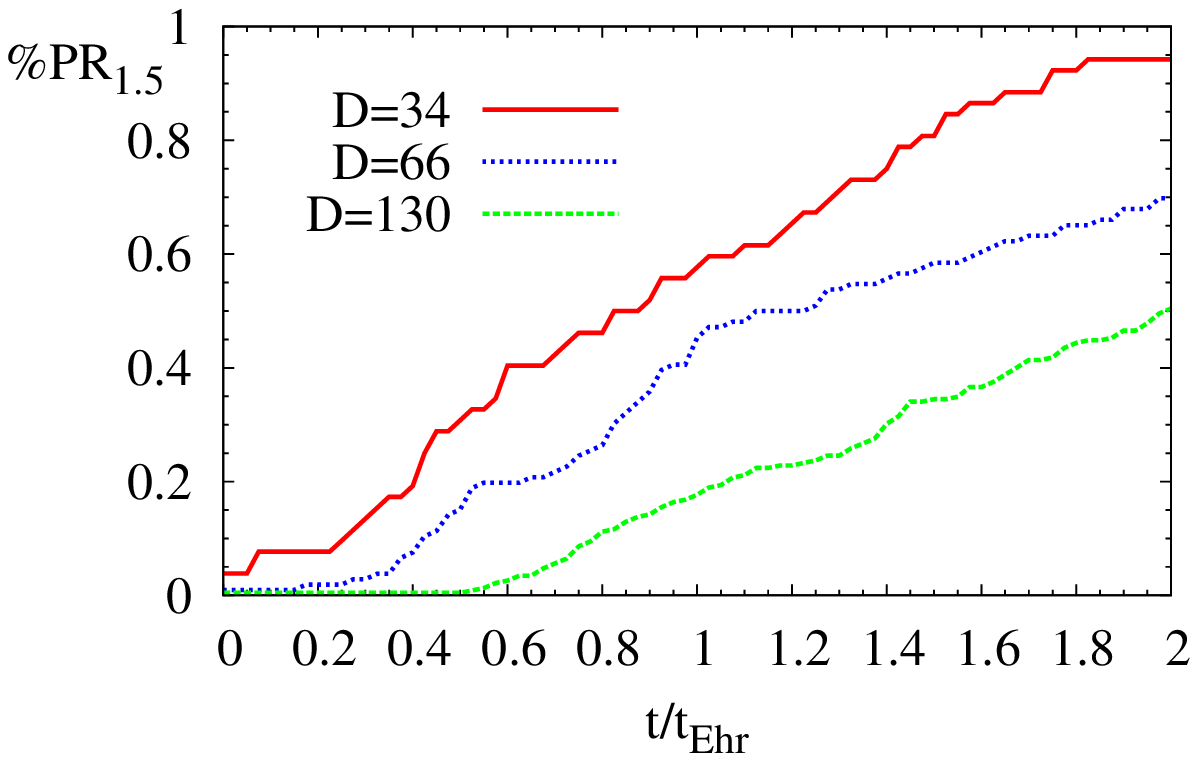}
\caption{(color online) Top: Average participation ratio ($\langle PR \rangle$) 
as a function of the propagation time for $D=34,66,130$ in units of Ehrenfest 
time; Bottom: $\%PR_{1.5}$, fraction of scar functions which have $PR<1.5$ in the QBM eigenstates 
basis.}\label{fig:PRBakdet}
\end{center}
\end{figure}  
\begin{figure}[tpb]
\begin{center}
 \includegraphics[width=0.48\textwidth]{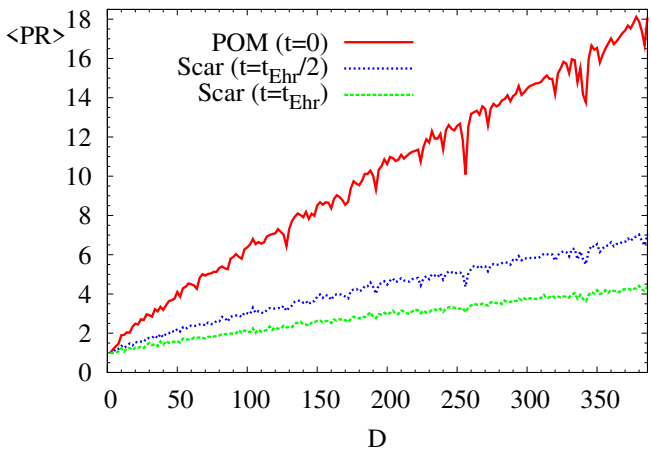}
\caption{(Color online) Average participation ratio ($\langle PR \rangle$) 
as a function of the dimension of the QBM for the POM and the scar 
functions with $t=t_{\text{Ehr}}/2$ and $t=t_{\text{Ehr}}$.}\label{fig:PRBakdeD}
\end{center}
\end{figure}

The scar functions have superposition with a small number of eigenstates.
This can be quantified in terms of the average participation ratio defined as
\begin{equation}
\langle PR\rangle=\frac{1}{D_{S}}\sum_{L=l}\left( \sum_{j=1}^{D}\vert\langle\psi_{j}\vert\Phi_{\bm{\nu}}^{t,k,R,T}\rangle\vert^{4}\right)^{-1}
\end{equation}
where $\langle PR \rangle\in[1,D)$. Notice that the random matrix theory prediction for a generic basis is  $\langle PR \rangle=D/2$.

The $\langle PR\rangle$ decreases with the propagation 
time and converges to $1$ in the Heisenberg time limit as was expected. 
However, the interesting region is for times of the order of the Ehrenfest time.
In Fig. \ref{fig:PRBakdet} (top) we show $\langle PR\rangle$ as a function of 
propagation time for $D=34,66,130$. In Fig. \ref{fig:PRBakdet} (bottom) we also show the fraction 
of scar functions which have $PR$ less than $1.5$ as a function of time.
Notice that for $D=130$ and at $t\simeq t_{Ehr}$ the average number of eigenstates in a scar is about $3$ and 
$20\%$ of the scar states have a $<PR>$ less than $1.5$ meaning that they are almost pure eigenstates.

The average participation ratio as a function 
of the dimension of the QBM is shown in Fig. \ref{fig:PRBakdeD} for the 
POM and the scar functions with $t=t_{\text{Ehr}}/2$ and $t=t_{\text{Ehr}}$.
For the limited range of values available the $\langle PR \rangle$ seems to grow linearly with $D$ 
but with a slope significantly smaller than the $D/2$ random matrix value. 
The reduction is similar (and more important) than that obtained in \cite{lakshmin,leo2}.

\section{Scar Approximation by Homoclinic Periodic Orbit Modes}\label{sec:homoc}

\begin{figure}[tpb]
\begin{center}
 \includegraphics[width=0.44\textwidth]{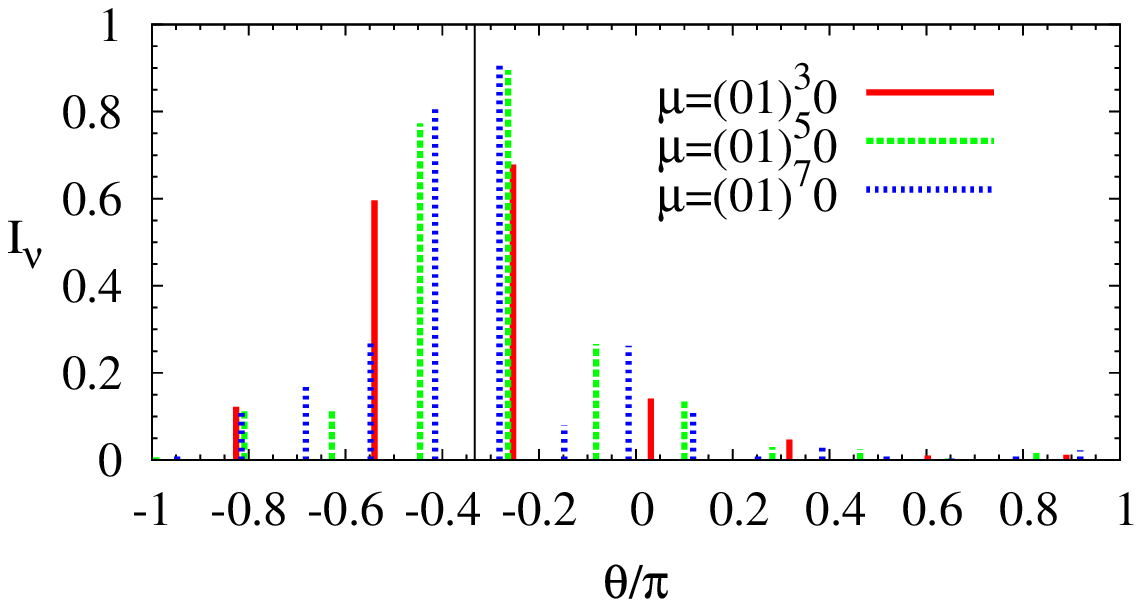}
 \includegraphics[width=0.44\textwidth]{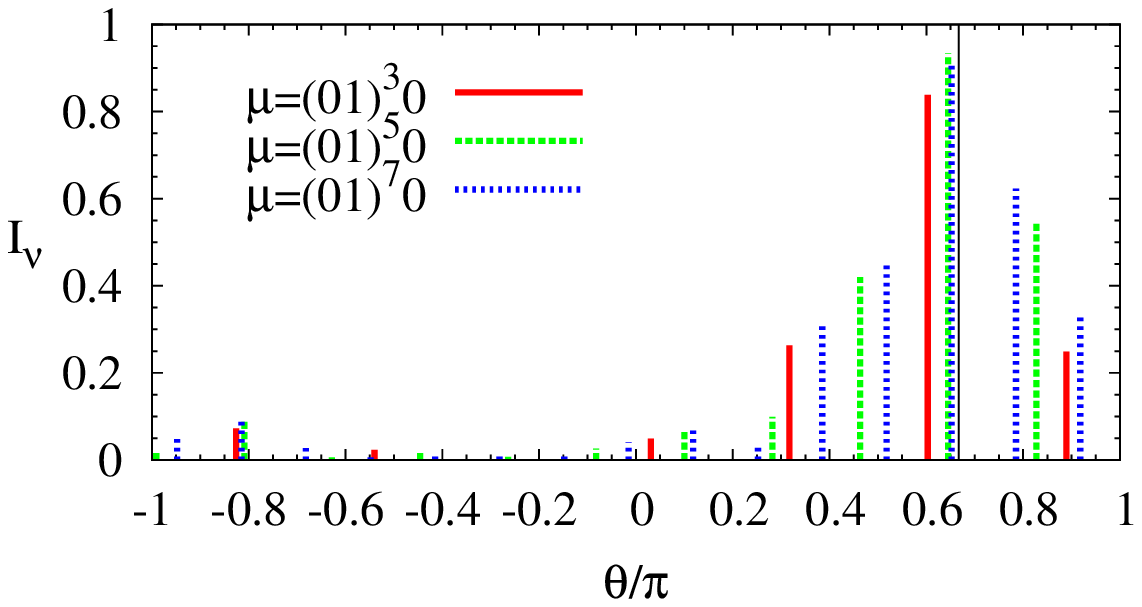}
\caption{(Color online) Intensity $I_{\bm{\nu}}\equiv\vert \langle \Phi_{\bm{\nu}}^{t_{\text{Ehr}/2},k_{\bm{\nu}}}\vert\Phi_{\bm{\mu}}^{0,k_{\bm{\mu}}}\rangle\vert$ 
between the scar function over $\bm{\nu}=01$ with $t=t_{\text{Ehr}}/2$, 
and the homoclinic POM over $\bm{\mu}=(01)^s0$ with $s=3,5,7$. $k_{\bm{\nu}}=0$ is shown on 
top and $k_{\bm{\nu}}=0$ on bottom, where respective phases are drawn with solid black lines. 
The dimension of the Hilbert space is $D=128$ and the products are ordered by increasing phase ($\theta$) 
of $\bm{\mu}$ for the $2s+1$ values of $k_{\bm{\mu}}$.
}\label{fig:scarvstube}
\end{center}
\end{figure} 
\begin{figure}[tpb]
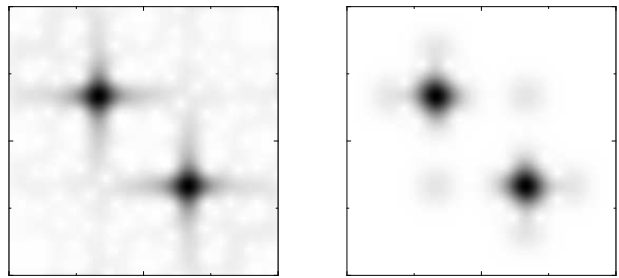

\begin{center}
 \includegraphics[width=0.2\textwidth,angle=90]{8a.ps}\hspace{0.8cm}
 \includegraphics[width=0.2\textwidth,angle=90]{8b.ps}
\caption{Husimi representation in the unit square phase space of the scar function $\vert\Phi_{\bm{\nu}}^{t_{\text{Ehr}/2},k=1}\rangle$ 
on the PO $\nu=01$ (left).
On the right, the Husimi representation of the POM $\vert \Phi_{\bm{\mu}}^{0,k=12}\rangle$ constructed on 
the homoclinic orbit $\bm{\mu}=(01)^s0$ (with $s=7$) which approximates the scar function. 
We plot the square of the Husimi function to enhance the structure on the manifolds. 
Those states are the scar function and the POM approximation of the scar of Fig. \ref{fig:scarvstube} respectively for $D=128$.
}\label{fig:scarvstubesHus}
\end{center}
\end{figure}

The construction of the POM basis is a simple analytical formula only involving as classical 
input short periodic orbits.
The scar function requires in addition the forward and backwards propagation of the POM.
In this paper for simplicity this propagation was obtained exactly by a matrix multiplication. 
It should be clear, however, that accurate semiclassical expressions for this propagation could 
be available as long as the times involved remain bounded by the Ehrenfest time.
In this section we illustrate a different approach that replaces the need for 
this propagation by the construction of POM's on long periodic orbits homoclinic to short ones.

%
The homoclinic POM are simply the POM construction described 
in Section \ref{sec:POM} over PO of the form 
$\bm{\mu}=(\bm{\nu})^s \bm{h}$ with period 
$L_{\bm{\mu}}=sL_{\bm{\nu}}+L_{\bm{h}}$. 
This PO resembles the homoclinic trajectories of $\bm{\nu}$ 
with the excursion $\bm{h}$ represented by the infinite string
$\ldots\bm{\nu\nu\nu h \nu\nu\nu}\ldots$, and converges to it in 
the long $s$ limit.
These orbits for large $s$ accumulate near stable and unstable manifolds of $\nu$ and therefore 
can mimic the building up of amplitude produced by the propagation.

A long POM with period $L_{\bm{\mu}}$ will generate $L_{\bm{\mu}}$ states with phases $A_{\bm{\mu}}^{k_{\bm{\mu}}}$
uniformly spaced on the unit circle. 
The state with the phase $A_{\bm{\mu}}^{k_{\bm{\mu}}}$ closer to $A_{\bm{\nu}}^{k_{\bm{\nu}}}$ of the short $\bm{\nu}$ orbit will have the 
largest overlap with the scar state. 
An example is shown in 
Fig. \ref{fig:scarvstube} for the product between the scar 
function on $\bm{\nu}=01$ for both values of $k_{\bm{\nu}}$ ($k=0$ on top and $k=1$ on bottom) with 
$t=t_{Ehr}$ and the homoclinic POM over $\bm{\mu}_s=(\bm{\nu})^s\bm{h}$ 
for all values of $k_{\bm{\mu}}$ ($I_{\bm{\nu}}\equiv\vert \langle 
\Phi_{\bm{\nu}}^{t_{\text{Ehr}},k_{\bm{\nu}}}\vert\Phi_{\bm{\mu}}^{t=0,k_{\bm{\mu}}}\rangle\vert$)
 as a function of the phase which take the discrete values $A_{\bm{\mu}}^k$. 
In this case we choose the shortest homoclinic excursion $\bm{h}=0$ 
and values of $s=3,5,7$. 
In this example acceptable approximations ($I_{\nu}\gtrsim 0.9$) are reached for values of $s$ no longer than $7$. 
In Fig. \ref{fig:scarvstubesHus} we compare the husimi representations of the scar function on $\bm{\nu}=01$ with 
$k=1$ and $t=t_{\text{Ehr}}/2$ and its best approximation with the homoclinic POM over $\bm{\mu}_s=(01)^s0$ with $s=7$. 
Note that the homoclinic POM structure looks similar to the evolution in time of the scar since it spreads 
the stable and unstable manifolds of the PO.

\section{Conclusions}

We have shown that a set of states constructed on the periodic orbits of the QBM provides a way 
of describing its eigenfunctions which significantly improves (in terms of the average participation ratio) 
the description in terms of a generic basis. For the QBM this improvement is similar to that obtained by Lakshminarayan 
using the Hadamard basis \cite{lakshmin} or by the essential baker map \cite{leo2}. 
However the method does not use any special structure of the chaotic map and should be generally applicable to other chaotic maps \cite{verginischneider2} as long as the properties of a limited set of periodic orbits are known.\\
The construction is still far from a full semiclassical analysis. 
The basis that we constructed is non--orthogonal and requires to consider each primitive orbit as non interacting with each other periodic orbit.
The semiclassical calculation of this interaction would be needed to calculate the amplitudes that describe the eigenstates in terms of scar states.
Calculation of this type has been performed for the billiard stadium \cite{vergini,vergini2,verginicarlo2}, an hyperbolic Hamiltonian \cite{verginischneider1}, and the cat map \cite{verginischneider2}.

It is also important to mention that the aim here is to reconstruct the \emph{unitary} dynamics of the map 
in terms of \emph{pure} states constructed as interpretations of gaussian packets with complex coefficients 
derived from the classical action. 
A different construction based on incoherent superposition of \emph{densities} placed on periodic points would 
allow a similar reconstruction in terms of the \emph{Liouville} dynamics. 
We postpone this aspect of the problem for future investigations.

We thank Eduardo Vergini for interesting discussions. Partial support by ANPCyT and PIP6137 of CONICET are
gratefully acknowledged.


  \renewcommand{\theequation}{A-\arabic{equation}}
  \setcounter{equation}{0}  
\section*{APPENDIX A: POM Diagonalization}\label{App}

\begin{figure}[htp]
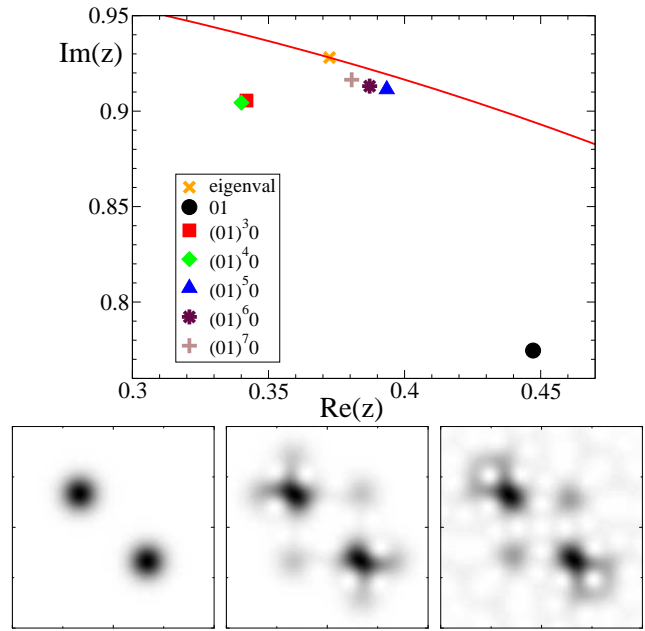

\begin{center}
 \includegraphics[width=0.4\textwidth]{9a.eps}\\
 \includegraphics[width=0.15\textwidth,angle=90]{9b.ps}\hspace{0.05cm}
 \includegraphics[width=0.15\textwidth,angle=90]{9c.ps}\hspace{0.05cm}
 \includegraphics[width=0.15\textwidth,angle=90]{9d.ps}
\caption{(color online) Top: Real and imaginary part of the solutions of the generalized eigenvalue problem for the 
PO $\bm{\nu}=01$ and the homoclinic orbit family $\bm{\mu}=(01)^s0$ for $s=3,4,5,6,7$. with $D=50$. 
These solutions converge to the exact eigenvalue which lives in the unit circle (in solid line). \\
Bottom: Husimi functions  in the unit square phase space of the states associated with the eigenvalues marked with $\bullet$ ($\bm{\mu}=01$),
 $+$ ($\bm{\mu}=(01)^70$) and $\times$ (the exact eigenstate of the map) on the left, center and right respectively.
}\label{fig:eigengral}
\end{center}
\end{figure}

The periodic orbit modes (POM) constructed in Section \ref{sec:POM} can also be seen in a different light.
Consider the generalized eigenvalue problem
\begin{equation}\label{eq:eigengral}
\det\left(\langle\gamma_i\vert\hat{B}\vert\gamma_j\rangle-z\langle\gamma_i\vert\gamma_j\rangle\right)=0
\end{equation}
where $\vert\gamma_i\rangle$ are coherent states placed on $M$ periodic points of the map.
The operator 
$ \sum_i^M\vert\gamma_i\rangle\langle\gamma_i\vert$
 can be considered as an approximation to the resolution of unity in the $D$--dimensional Hilbert space,
characteristic of coherent states as long as $M$ is large enough ($\gtrsim D$) and the points cover uniformly the phase space.
This latter condition is satisfied for chaotic maps on account of the Hannay--Ozorio de Almeida sum rule \cite{hannay}.
Under this conditions Eq. \ref{eq:eigengral} provides $D$ exact eigenvalues and $M-D$ zero eigenvalues \cite{haake}.

POM are obtained with the two assumptions:

a) Different primitive orbits do not interact.

b) Eq. \ref{eq:matrix} is satisfied.

With these conditions the eigenvalue problem is solved by a simple Fourier transform involving the different points on an orbit. The eigenfunctions are the POM's of Eq. \ref{eq:POM} and the eigenvalues are complex and given by Eq. \ref{eq:BSphase}.\\
If we retain only retain the assumption that periodic orbit do not interact we can refine the construction of POM by considering
 the limited diagonalization (Eq. \ref{eq:eigengral}) 
where now $\vert\gamma_i\rangle$ are the points on a primitive orbit (and its image under $R$ and $T$ if necessary). 
This refinement is important for long orbits of the type $\bm{\mu}=\bm{\nu}^s\bm{h}$ and $\bm{\mu}=\bm{\nu}_1^{s_{1}}\bm{h}\bm{\nu}_2^{s_2}$ which mimic homoclinic and heteroclinic orbits and accumulate 
 their points close to short primitive orbits.
For these orbits it is not correct to assume the coherent states as being even approximately orthogonal and therefore the generalized diagonalization is necessary.
The eigenvalues divide in two sets. Some converge rapidly to very small values, while there are always some of them which converge to the unit circle.\\
The advantage of this construction is that the ``good'' eigenvalues are closer to the unit circle and therefore represent modes that have a longer lifetime. Moreover spurious eigenvalues that are produced by the superposition of many almost equal coherent states are rapidly eliminated. The disadvantage is of course that the construction is not analytic and requires a diagonalization.\\ 
We give an example for the family of orbits $\bm{\mu}_s=\bm{\nu}^s\bm{h}$. We have diagonalized Eq. \ref{eq:eigengral} for periodic points on the orbit $\bm{\mu}_s$ ($s=3,4,5,6,7$) and compare with the exact eigenstate of the QBM for $D=50$.

\end{document}